\documentclass[conference]{IEEEtran}
\IEEEoverridecommandlockouts
\usepackage[letterpaper, left=0.625in, right=0.625in, bottom=1in, top=0.75in]{geometry}
\usepackage{cite}
\usepackage{ctable}
\usepackage{lipsum}
\usepackage{xspace}
\usepackage{subcaption}
\usepackage[font=small]{caption}
\usepackage[cmex10]{amsmath}
\usepackage{mathtools,amssymb,lipsum}
\usepackage{lettrine}
\usepackage{bm}
\usepackage{algorithm,algorithmic}
\usepackage{environ}
\NewEnviron{NORMAL}{%
    \scalebox{2}{$\BODY$} 
}
\NewEnviron{HUGE}{%
    \scalebox{5}{$\BODY$} 
} 
\usepackage{array}
\usepackage{url}  
\usepackage{amsthm}
\usepackage{multirow}
\usepackage{color}
\usepackage{epstopdf}
\usepackage{endnotes}
\usepackage{framed}
\usepackage{cool}
\usepackage{enumerate}
\usepackage{url}
\usepackage[subnum]{cases}
\usepackage[nocomma]{optidef}
\usepackage{etoolbox}
\usepackage{amsmath,amssymb,amsfonts}
\usepackage{graphicx}
\DeclareMathSizes{10}{9}{6}{5}
\usepackage{textcomp}
\usepackage{xcolor}
\def\BibTeX{{\rm B\kern-.05em{\sc i\kern-.025em b}\kern-.08em
    T\kern-.1667em\lower.7ex\hbox{E}\kern-.125emX}}
\begin{document}
\title{3TO: THz-Enabled Throughput and Trajectory Optimization of UAVs in 6G Networks by Proximal Policy Optimization Deep Reinforcement Learning\\
}
\author{\IEEEauthorblockN{$^\dagger$Sheikh Salman Hassan, $^\dagger$Yu Min Park, $^\dagger$Yan Kyaw Tun, $^\ddagger$Walid Saad, $^{\dagger\dagger}$Zhu Han, and $^\dagger$Choong Seon Hong}
\IEEEauthorblockA{$^\dagger$Department of Computer Science and Engineering, Kyung Hee University, Yongin, 17104, Republic of Korea\\
$^\ddagger$Wireless@VT, Bradley Department of Electrical and Computer Engineering, Virginia Tech, VA, 24061, USA\\
$^{\dagger\dagger}$Department of Electrical and Computer Engineering, University of Houston, Houston, TX 77004-4005, USA}
Email: \{salman0335, yumin0906, ykyawtun7, cshong\}@khu.ac.kr, walids@vt.edu, zhan2@uh.edu.
\thanks{This work was supported by the National Research Foundation of Korea (NRF) grant funded by the Korea government (MSIT) (No. No. 2020R1A4A1018607) and by the Institute of Information and Communications Technology Planning and Evaluation (IITP) Grant funded by the Korea Government (MSIT) (Artificial Intelligence Innovation Hub) under Grant 2021-0-02068. *Dr. CS Hong is the corresponding author.}
}
\maketitle
\begin{abstract}
Next-generation networks need to meet ubiquitous and high data-rate demand. Therefore, this paper considers the throughput and trajectory optimization of terahertz (THz)-enabled unmanned aerial vehicles (UAVs) in the sixth-generation (6G) communication networks. In the considered scenario, multiple UAVs must provide on-demand terabits per second (TB/s) services to an urban area along with existing terrestrial networks. However, THz-empowered UAVs pose some new constraints, e.g., dynamic THz-channel conditions for ground users (GUs) association and UAV trajectory optimization to fulfill GU's throughput demands. Thus, a framework is proposed to address these challenges, where a joint UAVs-GUs association, transmit power, and the trajectory optimization problem is studied. The formulated problem is mixed-integer non-linear programming (MINLP), which is NP-hard to solve. Consequently, an iterative algorithm is proposed to solve three sub-problems iteratively, i.e., UAVs-GUs association, transmit power, and trajectory optimization. Simulation results demonstrate that the proposed algorithm increased the throughput by up to 10$\%$, 68.9$\%$, and 69.1$\%$  respectively compared to baseline algorithms.\\
\end{abstract}
\section{Introduction}
Wireless communication systems have grown exponentially in the past several decades due to the need for high-speed data connections wherever and whenever needed. Therefore, sixth-generation (6G) networks must be multi-dimensional and capable of providing ubiquitous and diverse services by integrating current terrestrial network specifications with space and air-based information networks \cite{Prof_walid_6G, own_globecom, hassan2022seamless}. 6G networks will likely be cell-free, which means that users will be able to smoothly and automatically switch from one network to another to seek the most appropriate and qualified communication without the need for human administration and settings \cite{6g+THz_survey}. 

Furthermore, the terahertz (THz) band, with frequencies ranging from 0.1 to 10 THz, is a viable frequency range for the next generation of ultra-dense wireless networks \cite{6g+THz_survey2, chaccour2021terahertz}. The THz channel can provide rates in the order of terabits per second (Tbps) that are suitable to support emerging applications, including high-quality video streaming, virtual and augmented reality, and chip-based wireless networks. Moreover, due to the restricted bandwidth at millimeter-wave (mmW), i.e., 30 gigahertz (GHz) carrier frequencies, it is impossible to reach Tbps data speeds. 

In particular, utilizing THz-enabled unmanned aerial vehicles (UAVs) communications to offer seamless coverage and provide high bandwidth to ground users could be an efficient wireless network integration. Many mobile carriers in the United States have tested UAV-mounted LTE base stations, including AT\&T and Verizon \cite{ATTANDVERIZON}. Aerial base stations may be attached to UAVs at low altitudes, making them more cost-effective, quick, and adaptable than terrestrial communication infrastructures \cite{6G_salman}. Also, UAV communications benefit from better line-of-sight links with ground users due to their high altitudes. However, there are significant problems in terms of deployment, trajectory design, and network resource optimization when using UAVs for wireless communications.

To fully exploit the design degrees of freedom for THz-enabled UAV communications, it is crucial to investigate the UAV's mobility and its network resource management in three-dimensional space. The existing works \cite{uav_ref1, UAV_ref2, UAV_ref3} consider the UAVs as aerial base stations. The authors in \cite{uav_ref1} studied the problem of minimization of uplink power through user association in the presence of a single UAV base station. In \cite{UAV_ref2}, the authors investigated the problems of optimal trajectory design and transmit power optimization to maximize data rate in the presence of energy harvesting constraints. Similarly, the authors in \cite{UAV_ref3} studied a three-dimensional (3D) coverage maximization problem for UAV networks. Moreover, the authors in \cite{UAV_THz_ref1} analyzed the coverage probability of UAVs with THz communication but did not optimize the UAVs' deployment. The authors in \cite{THz_channel_ref} and \cite{UAV_comm} investigated the problem of UAV deployment and transmitting power for THz communication. However, they used traditional optimization techniques to solve the proposed problem. Moreover, none of the prior research takes into account the usage of the THz spectrum for UAVs with deployment and trajectory optimization.

Hence, to fill the knowledge gap, we investigate the fundamental study of optimal UAV trajectory deployment design and network resource management at THz frequencies. Nonetheless, given the significant degree of uncertainty in higher frequency bands such as THz, it is critical to provide additional degrees of freedom and control in network management. Therefore, given the ability of UAVs to fly, they are suitable for providing line-of-sight (LoS) communication links to ground users (GUs) \cite{Prof_Zhu1} and \cite{Prof_Zhu2}. To reap the benefits of deploying THz-enabled UAVs, it is important to optimize the locations of the UAVs to offer continuous LoS connections to GUs. Therefore, to handle the mobility of UAVs in our proposed dynamic network, we consider deep reinforcement learning-based UAVs location optimization with a low-complexity algorithm for GU association and transmit power optimization. The key contributions of our proposed work can be summarized as follows:
\begin{itemize}
    \item We propose a THz-enabled UAVs communication network architecture by considering the quality-of-service (QoS) parameters for the GUs and the UAV's optimal trajectory deployment.
    \item Our objective is to maximize overall throughput between the UAVs and the GU by jointly optimizing the operational UAV's trajectory deployment and GU association, as well as minimizing the transmitting power of the UAVs. 
    \item To tackle this optimization problem, we propose an iterative algorithm that separates the original optimization problem into three subproblems: A GU association subproblem is handled by balanced K-means clustering (BKMC), a power control subproblem is solved by successive convex approximation (SCA), and a trajectory deployment optimization subproblem is tackled by proximal policy optimization deep reinforcement learning (PPO-DRL) that is solved iteratively.
    \item Numerical results show that the proposed algorithm can increase throughput by up to 10$\%$, 68.9$\%$, and 69.1$\%$ respectively when compared to a baseline that optimizes only transmit power with static UAVs, random transmit power with static UAVs and optimizes only UAVs trajectories with random transmit power.
\end{itemize}

The rest of the paper is organized as follows. The system model and the problem formulation are given in Section \ref{system_model} and \ref{prob_form} respectively. Then, the proposed algorithm is presented in Section \ref{sol_algo}. Section \ref{simul} describes the implementation and simulation results, and Section \ref{conclusion} concludes the paper. 

\section{System Model \& Problem Formulation}
\subsection{Network Model}
\label{system_model}
\begin{figure}[t!]
    \centering
    \includegraphics[width=\columnwidth]{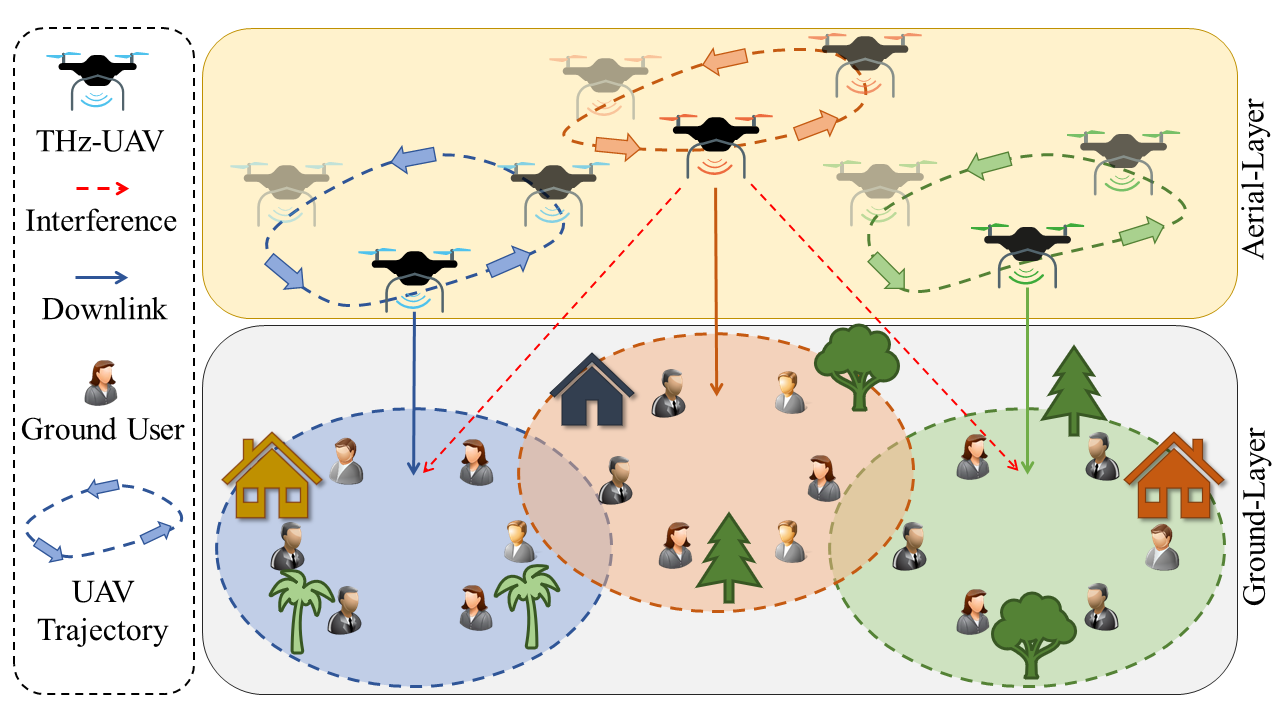}
    \caption{Illustration of THz-enabled UAVs Network.}
    \label{sysmod}
\end{figure}

As shown in Fig. \ref{sysmod}, we consider a THz-enabled multi-UAV wireless communication network. This system model consists of a set $\mathcal{K}$ of $K$ UAVs that seek to provide communication services to a set $\mathcal{M}$ of $M$ GUs distributed according to a homogeneous Poisson point process (HPPP). The three-dimensional (3D) coordinates of UAV $k$ can be represented as $\boldsymbol{q_{k}}(n)$=$\big(x_{k}(n), y_{k}(n), z_{k}(n)\big)$, and the two-dimensional (2D) coordinates of GU $m$ can defined as $\boldsymbol{o_{m}}(n)$=$\big(x_{m}(n), y_{m}(n)\big)$ respectively. 
\subsection{Channel Model \& Link Analysis}
The presence of this molecular absorption loss distinguishes the terahertz channel. This loss is produced by molecules in the air, such as $\mathrm{H}_2\mathrm{O}$ vapor, which each have their own absorption spectrum, making the wireless channel frequency selective. Thus, the THz communication channel between UAV $k$ and GU $m$ can be modeled as follows \cite{THz_channel_ref}:
\begin{equation}
    h_{k,m} = d^{-2}_{k,m}e^{-a(f)d_{k,m}}, \quad \forall k \in \mathcal{K},~\forall m \in \mathcal{M},
\end{equation}
where $d_{k,m} = \|\boldsymbol{q}_{k} - \boldsymbol{o}_{m}\|$ is the distance between UAV $k$ and GU $m$, $e^{-a(f)d_{k,m}}$ represents the channel path-loss due to the molecular absorption, and $a(f)$ denotes the molecular absorption coefficient, which depends on the network operating frequency (i.e., THz frequency) and the concentration of water vapor molecules in the air\footnote{Hereinafter, for the sake of simplicity, we will write $a$ instead of $a(f)$.}.
To efficiently use the network resources, we consider each UAV to utilize the same frequency spectrum. Therefore, the network experiences interference from non-associated UAVs to the GUs in each time slot, which can be defined as:
\begin{equation}
    \psi_{k,m}(n) = \sum_{\forall k' \ne k} \sum_{\forall m' \ne m} {p_{k',m'}(n) h_{k',m'}(n)}, \quad \forall k \in \mathcal{K},~\forall m \in \mathcal{M},
\end{equation}
where $p_{k',m'}$ is the transmit power, $k'$ is the non-associated UAVs, and $m'$ represents the non-associated GUs, respectively. Therefore, the signal-to-interference-plus-noise ratio (SINR) from each UAV $k$ to GU $m$ in each time slot $n$ can be formulated as:
\begin{equation}
    \gamma_{k,m}(n) = {{p_{k,m}(n) h_0} \over {\psi_{k,m} + \alpha_{k,m}B_{k,m} {d^{2}_{k,m}} e^{ad_{k,m}} \sigma^2}}, \quad \forall k \in \mathcal{K},~m \in \mathcal{M},
\end{equation}
where $h_0$ is the channel gain at a reference distance $d_0=1$m, $p_{k,m}$ is the transmit power from the UAV $k$ to their associated GU $m$, $\alpha_{k,m}$ proportion of the channel bandwidth from the UAV $k$ to GU $m$, $B_{k,m}$ is the total channel bandwidth, and $\sigma^2$ is the additive white Gaussian noise (AWGN) power. The achievable throughput from UAV $k$ to GU $m$ in each time slot $n$ can be obtained based on the SINR $\gamma_{k,m}$ as:
\begin{equation} 
\label{eq1}
R_{k,m} (n) = \alpha_{k,m} B_{k,m} \log_2 \left( 1 + \gamma_{k,m} \right).
\end{equation}
\subsection{Problem Formulation}
\label{prob_form}
Our goal is to maximize the total throughput from all the deployed UAVs while satisfying the QoS and trajectory constraints of each GU and UAV, respectively, in the network. Therefore, the throughput maximization problem can be defined as:
\begin{maxi!}|s|[2]
		{\substack{\boldsymbol{\alpha, p, q}}}
		{ R^{\textrm{lo}}_{k}(n) \label{obj1}}{\label{opt:P}}{\textbf{P1:}}
		\addConstraint{ \sum_{n=1}^{N} \sum_{m=1}^{M} \alpha_{k,m} R_{k,m}(n) \geq R^{\textrm{lo}}_{k}(n), ~\forall k \in \mathcal{K},{\label{c1}}}
		\addConstraint{\alpha_{k,m} R_{k,m}(n) \geq R^{\textrm{min}},~\forall m \in \mathcal{M}, k \in \mathcal{K}, n \in \mathcal{N}, {\label{c2}}}
		\addConstraint{ \alpha_{k,m} \in \left\{0,1 \right\}, \sum_{m=1}^{M} \alpha_{k,m} = 1, ~\forall k \in \mathcal{K}, m \in \mathcal{M}, {\label{c3}}}
		\addConstraint{ \sum_{m=1}^{M} p_{k,m}(n) \leq P^{\textrm{max}}_{k}, \forall k \in \mathcal{K}, n \in \mathcal{N}, {\label{c4}}}
		\addConstraint{ 0 \leq p_{k,m}(n) \leq P^{\textrm{max}}_{k} ,\forall k \in \mathcal{K}, n \in \mathcal{N}, {\label{c5}}}
		\addConstraint{\lVert \mathbf{q}_{i}(n) - \mathbf{q}_{j}(n) \rVert^2_2 \geq D^2_{\mathbf{min}} ,   \forall i\neq j \in \mathcal{K} ,   \forall n \in \mathcal{N} {\label{c6}}},
		\addConstraint{{\lVert \mathbf{q}_k(n+1)-\mathbf{q}_k(n) \rVert \over{ t^{\textrm{mov}}}} \leq V^\mathbf{max},   \forall k \in \mathcal{K}, n \in \mathcal{N}{\label{c7}} },
\end{maxi!}
where ({\ref{c1}}) represents the optimization objective, which is the sum-rate of each UAV at each time slot $n$. ({\ref{c2}}) ensures the QoS constraint of each user from the associated UAV, ({\ref{c3}}) presents that each GU can be associated with at most one UAV, ({\ref{c4}}) and ({\ref{c5}}) ensures the total transmit power of the UAV have to be less than the maximum transmit power $P^{\textrm{max}}_{k}$ of the UAV, ({\ref{c6}}) guarantees that the distance between UAVs is not as close as the minimum distance $D_{\textrm{min}}$, and ({\ref{c7}}) is the UAVs speed constraint.

\section{Proposed Algorithm}
\label{sol_algo}
$\boldsymbol{\mathrm{P1}}$ is a MINLP, which is NP-hard and difficult to solve. Therefore, we will decompose it into three subproblems and propose an iterative algorithm that is composed of BKMC, SCA, and PPO-DRL in the following subsections, respectively.
\subsection{Balanced K-means Clustering}
Given the initial UAVs' transmit power $\boldsymbol{p}$ and trajectory $\boldsymbol{q_m}$, $\boldsymbol{\mathrm{P1}}$ is transformed into $\boldsymbol{\mathrm{P1.1}}$ for optimizing GU association $\boldsymbol{\alpha}$, which is integer linear programming. Mathematically, it can be defined as:
\begin{maxi!}|s|[2]
		{\substack{\boldsymbol{\alpha}}}
		{R^{\textrm{lo}}_{k}(n) }{\label{opt:P1.1}}{\textbf{P1:1}}
		\addConstraint{({\ref{c1}}), ({\ref{c2}}),\ \hbox{and} \ ({\ref{c3}) }.}
\end{maxi!}
GU association problems, in general, can be solved by K-means clustering. Although K-means clustering poses an unfairness issue that can lead to any cluster expanding excessively. Thus, we propose the use of BKMC, which can calculate the size of each cluster in advance \cite{malinen2014balanced}.

We follow the same steps as K-means, while the assignment phase is different. We define pre-allocated slots according to the total number of GU $M$, instead of choosing the closest UAV $k$, each GU $m$ needs to be associated with these slots i.e., $M/K$ slots per cluster. Assuming that $\lfloor M/K \rfloor=\lceil M/K \rceil = M/K$, this will compel all clusters to have the same size. Otherwise, there will be ($M$ mod $K$) clusters of size $\lfloor M/K \rfloor$, and $K-$($M$ mod $K$) clusters of size $\lceil M/K \rceil$. To find an assignment that minimizes mean square error (MSE), we solve an assignment problem using the Hungarian algorithm \cite{Hungarian}.

In contrast to a traditional assignment problem with fixed weights, the weights in this problem vary dynamically after each K-means iteration based on the newly determined centroids. The Hungarian method is then used to obtain the minimum weight pairing. The update step for choosing new UAV locations is similar to normal K-means centroids, where the new location are calculated as the means of each GU location: $\boldsymbol{q}_m$ assigned to each cluster $i$:
\begin{equation}
C^{(t+1)}_{i} = {1 \over n_{i}} \sum_{\boldsymbol{o}^i_{m} \in C^{(t)}_{i}} \boldsymbol{o}^i_{m},
\end{equation}
where $\boldsymbol{o}^i_{m}$ represents GU $m$ location in cluster $i$ and $C_i$ denotes the UAV (centroid) location. The edge weight is the distance between the UAV and GU, which is updated following the update step in each iteration. The description of BKMC is given in Algorithm \ref{alg:BKMC}.
\subsection{Successive Convex Approximation}
Given the optimal GUs association $\boldsymbol{\alpha}^*$ and UAVs trajectory $\boldsymbol{q_m}$, $\boldsymbol{\mathrm{P1}}$ is transformed into $\boldsymbol{\mathrm{P1.2a}}$ for optimizing transmit power $\boldsymbol{p}$, which is non-convex programming. Mathematically, it can be defined as:
\begin{algorithm}[t]
	\caption{\strut BKMC for GUs Association} 
	\label{alg:BKMC}
	\begin{algorithmic}[1]
	    \STATE{\textbf{Input:} the GU locations $\{\mathbf{o}_m\}_{m \in \mathcal{M}}$, the initial UAV locations $\{\mathbf{q}_k\}_{k \in \mathcal{K}}$}. 
		\STATE{\textbf{Initialize:} Initialize centroid locations $C^0$ to UAV locations $\{\mathbf{q}_k\}_{k \in \mathcal{K}}$}.
		\STATE{$t \leftarrow 0$}
		\REPEAT
		\STATE{Calculate distances between GUs and UAVs}.
		\STATE{Solve an assignment problem by Hungarian algorithm}.
		\STATE{Calculate new centroid locations $C^{t+1}$}.
		\UNTIL{the positions of the centroids do not change}
		\STATE{\textbf{Output:} Optimal user association. $\boldsymbol{\alpha}^*$}
	\end{algorithmic}
	\label{Algorithm}
\end{algorithm}
\begin{maxi!}|s|[2]
		{\substack{\boldsymbol{{p}}}}
		{R^{\textrm{lo}}_{k}(n) }{\label{P1.2}}{\textbf{P1.2a:}}
		\addConstraint{({\ref{c1}}), ({\ref{c4}), \ \hbox{and} \ ({\ref{c5}})  }.}
\end{maxi!}
To transform the non-convex objective, we apply SCA, which is based on the  first-order Taylor approximation, knowing that this provides the global upper bound for the concave function. To develop the SCA-based algorithm, we rewrite the log-term in objective (\ref{obj1}) in the form of the difference of two concave (D.C.) functions, which can be defined as:
\begin{equation}
    R(\boldsymbol{p}) = \sum^{K}_{k=1} R^{lo}_{k} = \log_2 \left( 1 + \gamma_{k,m} \right) = l(\boldsymbol{p})-h(\boldsymbol{p}),  \label{DC} 
\end{equation}
where
\begin{equation}
\label{dc1}
    l(\boldsymbol{p}) = \sum^{K}_{k=1} \sum^{M}_{m=1} \log_2 \left({p_{k,m}(n)h_0} \right),
\end{equation}
and
\begin{equation}
\label{dc2}
  h(\boldsymbol{p}) = \sum^{K}_{k=1} \sum^{M}_{m=1} \log_2 \left( {\psi_{k,m} + B_{k,m} {d^2_{k,m}(n)} e^{a d_{k,m}} \sigma^2} \right).
\end{equation}
The functions in (\ref{dc1}) and (\ref{dc2}) are concave, but the difference between them, captured in (\ref{DC}), is neither convex nor concave. Therefore, we presented the SCA, which can calculate the concave lower bound, i.e., the surrogate function for the non-concave objective given in (\ref{DC}), by providing a feasible solution $\boldsymbol{p'}$ of the problem (\ref{opt:P}). We design its lower bound with substituting $h(\boldsymbol{p})$ by their first-order Taylor approximation which can be defined as:
\begin{equation}
\label{first_order_expan}
  R(\boldsymbol{p}, \boldsymbol{p'}) = l(\boldsymbol{p}) - \Tilde{h}((\boldsymbol{p}, \boldsymbol{p'})),
\end{equation}
where 
\begin{equation}
 \label{taylor_exp}
   \Tilde{h}((\boldsymbol{p}, \boldsymbol{p'}))  \triangleq  h(\boldsymbol{p'}) - \nabla h(\boldsymbol{p'})(\boldsymbol{p} - \boldsymbol{p'}),
\end{equation}
where (\ref{taylor_exp}) is the first-order Taylor's expansion of $h(\boldsymbol{p})$ near the given point $\boldsymbol{p'}$ in the feasible region of the solution space, and $\nabla h(\boldsymbol{p'})$ denotes the gradient of the $h(\boldsymbol{p})$ at $\boldsymbol{p'}$. The gradient for UAV $k$ can be given as:
\begin{equation}
\begin{aligned}
    \nabla_k h(\boldsymbol{p'})&= \frac{\partial h(\boldsymbol{p'})}{\partial p'_k} 
    = \frac{1}{\mathrm{ln} 2}\\& \sum_{\forall k' \ne k} \sum_{\forall m' \ne m} \frac{h_0}{\psi_{k,m} + \alpha_{k,m}B_{k,m} {d^2_{k,m}} e^{-ad_{k,m}{n}} \sigma^2}. 
\end{aligned}
\end{equation}
The surrogate function is concave  given in (\ref{taylor_exp}). Additionally, we can find the upper bound of function $h(\boldsymbol{p})$ by the first-order Taylor's expansion as:
\begin{equation}
\label{upper_bound}
    h(\boldsymbol{p}) \leq h(\boldsymbol{p'}) + \nabla h(\boldsymbol{p'})(\boldsymbol{p} - \boldsymbol{p'}).
\end{equation}
By analysing (\ref{DC}), (\ref{first_order_expan}), and (\ref{upper_bound}), we can deduce the following observations:
\begin{equation}
\label{aprox}
    \begin{aligned}
        R(\boldsymbol{p}) & = l(\boldsymbol{p})-h(\boldsymbol{p})\\
        & \geq l(\boldsymbol{p}) - \{h(\boldsymbol{p'})+ \nabla h(\boldsymbol{p'})(\boldsymbol{p} - \boldsymbol{p'}) \}\\
        &\geq l(\boldsymbol{p}) - h(\boldsymbol{p'}) - \nabla h(\boldsymbol{p'})(\boldsymbol{p} - \boldsymbol{p'})\\
        & = R(\boldsymbol{p}, \boldsymbol{p'}).
    \end{aligned}
\end{equation}
where ($\ref{aprox}$) represents that the surrogate function provide the lower bound of the original function. Therefore, these two functions are tangent at point $\boldsymbol{p'}$, i.e., $R(\boldsymbol{p}, \boldsymbol{p'})|_{p=p'}$=$R(\boldsymbol{p'})$. Thus, the function in ($\ref{aprox}$) can provide the lower bound for the original objective function in ($\ref{opt:P}$). Therefore, we replace the non-concave objective in ($\ref{opt:P}$) with its surrogate function given ($\ref{first_order_expan}$). Afterward, we modify the surrogate objective function and found the convex problem which can be given as:
\begin{mini!}|s|[2]
		{\substack{\boldsymbol{{p}}}}
		{-R^{\textrm{lo} }_{k}(n) }{\label{opt:P1.2b}}{\textbf{P1.2b:}}
		\addConstraint{({\ref{c1}}), ({\ref{c4}), \ \hbox{and} \ ({\ref{c5}})  }.}
\end{mini!}
Hence, problem ($\ref{opt:P1.2b}$) is convex, and we can solve it using optimization solvers, i.e., CVXPY. We observe that the optimal solution $\boldsymbol{p^*}$ in each iteration provides the new reference point for $\boldsymbol{p'}$. By updating the value of $\boldsymbol{p'}$, we can define a new surrogate function and find the new approximated convex problem. This procedure will converge iteratively till the convergence criteria value $\chi$ is met. The details of SCA-approximation are shown in Algorithm $\ref{alg:SCA}$. Step 3 represents the convex approximation and step 4 obtains the successive update based on the newly obtained solution respectively in Algorithm $\ref{alg:SCA}$.
\begin{algorithm}[t]
	\caption{\strut SCA for Transmit Power Optimization ($\ref{opt:P1.2b}$)} 
	\label{alg:SCA}
	\begin{algorithmic}[1]
	    \STATE{\textbf{Input:} $p^{\mathrm{max}}_{k,m}$, $\boldsymbol{p}^0$, iteration~$j=0$, tolerance~$\chi$, stopping criterion~$e = 1$}.
		\STATE{$j \leftarrow 0$}
		\WHILE {e $\geq$ $\chi$}
		\STATE{Designed $R(\boldsymbol{p}, \boldsymbol{p'}) = l(\boldsymbol{p}) - \Tilde{h}((\boldsymbol{p}, \boldsymbol{p'}))$ based on ($\ref{first_order_expan}$)}.
		\STATE{Solve ($\ref{opt:P1.2b}$) and find the $\boldsymbol{p}^{j+1}$}.
		\STATE{Calculate the stopping criterion $e = |R(\boldsymbol{p}^{j+1})-R(\boldsymbol{p}^j)|$}.
		\STATE{Update the iteration counter i.e., $j=j+1$}.
		\ENDWHILE
		\STATE{\textbf{Output:} Optimal transmit power $\boldsymbol{p}^*$}.
	\end{algorithmic}
	\label{Algorithm}
\end{algorithm}
\subsection{Proximal Policy Optimization}
The UAV trajectory optimization $\boldsymbol{q}$ problem is still non-convex by giving the optimal user association $\boldsymbol{\alpha}^*$ and transmit power $\boldsymbol{p}^*$.
Mathematically, this problem can be defined as:
\begin{maxi!}|s|[2]
		{\substack{\boldsymbol{{q}}}}
		{R^{\textrm{lo}}_{k}(n) }{\label{opt:D}}{\textbf{P1.3:}}
		\addConstraint{({\ref{c1}}), ({\ref{c2}}), ({\ref{c6}}), \ \mbox{and} \ ({\ref{c7}}).}
\end{maxi!}
To address the challenge of dynamic UAV trajectory optimization, we propose a PPO-based DRL. PPO-DRL is based on the substitution of flexible constraints for hard constraints, which are seen as penalties. The new, more manageable constraints are utilized to solve a first-order differential problem that approximates the second-order optimization differential equation. The Kullback–Leibler (KL) divergence is used to calculate policy changes at each time step \cite{schulman2017proximal}. PPO seeks to compute an update that minimizes the cost function while keeping the divergence from the prior policy to a minimum at each time step. Trust region policy optimization (TRPO) recommends optimizing surrogate loss and limiting the amount of the update using KL. The PPO objective implements a method for updating the trust region that is consistent with stochastic gradient descent (SGD) and simplifies the algorithm by eliminating the KL penalty and the necessity for adaptive updates. Thus, PPO offers the advantages of being simple to apply and having a reduced level of complexity, which is appropriate for our highly dynamic environment. The surrogate objective function in TRPO can be maximized subject to a size limit on the policy update, which can be described as \cite{schulman2015trust}:
\begin{equation}
L^{\textrm{TRPO}}(\theta)= \hat{E}_{t} \left[ {\pi_{\theta}(a_{t}|s_{t}) \over \pi_{\textrm{old}}(a_{t}|s_{t})} \hat{A}_{t} \right] = \hat{E}_{t} \left[ r_{t}(\theta) \hat{A}_{t} \right], \label{L_CPI}
\end{equation}
where $\pi_{\theta}$ is a stochastic policy, $\hat{A}_{t}$ is an estimator of the advantage function at step $t$, $\theta_{\textrm{old}}$ is the vector of policy parameters before the update, $r_{t}(\theta)$ denotes the probability ratio, which can be define as:
\begin{equation}
    r_{t}(\theta)= {\pi_{\theta}(a_{t}|s_{t}) \over \pi_{\textrm{old}}(a_{t}|s_{t})},
\end{equation}
and $\hat{E}_{t}[...]$ indicates an expectation of the empirical average over a finite batch of samples in the algorithm that alternates between sampling and optimization. At this point, learning fails or performance suffers as a result of an excessive rise in the probability ratio $r_{t}(\theta)$, and in the case of TRPO, the problem is avoided by utilizing KL-Divergence to impose a penalty. However, TRPO has the drawback of being conceptually difficult to grasp, and hence difficult to apply. PPO, like TRPO, optimizes the surrogate objective function via stochastic gradient ascent (SGA). Thus, PPO used computationally efficient penalties and avoided unnecessary policy changes by employing a technique known as the clipped surrogate objective. The clip is a function that determines the minimum and maximum values of a given variable. The probability ratio $r_{t}(\theta)$ can be trimmed by PPO. The following is the objective function to which the clipping is applied:
\begin{equation}
L^{\textrm{PPO}}(\theta)=\hat{E}_{t} \left[ \textrm{min}(r_{t}(\theta)\hat{A}_{t}, \textrm{clip}(r_{t}(\theta), 1-\epsilon, 1+\epsilon)\hat{A}_{t}) \right],\label{L_CLIP}
\end{equation}
where $\epsilon$ is a hyper parameters. The function in \eqref{L_CLIP} compares the objectives used in the existing TRPO with the objectives to which clipping is employed and takes a smaller value.

Another feature of the PPO method is that, unlike many other DRL algorithms, the PPO-Actor-Critic networks contain a type of parameter sharing, which allows each objective to be combined into a single goal function and optimized at the same time. The methods for power allocation mentioned above are added to the function for computing rewards once the actor-critic network has been updated and learned. With this, it was possible to effectively separate power allocation and UAV trajectory and proceed with optimization. Next, the state, compensation, and behavior of the agent used in learning will be described. In learning step $t$, the state $s_{t}(n)$ is defined as:
\begin{equation}
s_{t}(n) = \{ \{\mathbf{q}_{k}(n)\}_{k\in \mathcal{K}}, \{\mathbf{o}_m\}_{m\in \mathcal{M}} \},\label{eq1}
\end{equation}
where $\mathbf{q}_{k}(n)$ and $\mathbf{o}_m(n)$ denotes UAV $k$ and GU $m$ locations, respectively, at time slot $n$. Therefore, agent takes optimal action through network with all location information input without any other information. The action in learning step $t$ at time slot $n$ is the speed and the moving direction as follow:
\begin{equation}
a_{t}(n)=\{ \{ v_k(n), \phi_k(n)\}_{k \in \mathcal{K}} \}\label{eq1}.
\end{equation}
Specifically, we restricted the direction $\phi_k(n)$ between $\left[ -\pi/3, \pi/3 \right]$ to prevent UAVs from turning sharply. Lastly, the reward in learning step $t$ at time slot $n$ is divided into three and can be expressed as follows:
\begin{equation}
r_{t}(n)=
\begin{cases}
2, & \mbox{if }t=\textrm{max step}, \\
-2, & \mbox{if }  \exists ~i, j \in \mathcal{K}\\
&\mathrm{ s.t.}~\lVert \mathbf{q}_{i}(n) - \mathbf{q}_{j}(n) \rVert < D_{\mathbf{min}},\\
\sum_{k=1}^{K} R^{\textrm{lo}}_{k}(n), & \mbox{otherwise}.
\end{cases}
\label{reward}\end{equation}
The episode finishes when learning step $t$ reaches the maximum step with a reward of $2$ and if the distance between UAVs is as near to the minimum distance then the penalty will be $-2$. Otherwise, it receives the total of the minimum data rate given by each UAV with the proposed power allocation and GU association algorithm. 

The pseudocode for the PPO learning process for UAV trajectory is depicted in Algorithm \ref{alg:PPO}. We proceed with the learning of the $E$ number of episodes. The episode begins by randomly arranging the locations of GUs. Algorithm 1 determines the optimal GU association $\boldsymbol{{\alpha}}^*$ based on the location of the arranged GUs. In subsequent learning, $A$ actors simultaneously generate data on the environment. Each actor acts under prior policy $\pi_{\theta_{\textrm{old}}}$ and employs Algorithm 2 to determine optimal power $\boldsymbol{{p}}^*$ and derive rewards from it. Trajectory memory stores the state, action, reward, and future state of time step $n$. When the time step is completed, we calculate the advantage estimates $\hat{A}_{n}$ for each time step $n$. Trajectory memory is used to extract minibatch, which is then optimized for the target function $L^{\textrm{PPO}}$. The learning process is repeated by substituting the updated parameter $\theta$ with the old value $\theta_{\textrm{old}}$. As a result, we can obtain a PPO network $\pi_{\theta_{\textrm{opt}}}$ for UAVs trajectory.

\subsection{Algorithms Complexities and Convergences}
Based on the definition mentioned above, the detailed process of resource allocation via BKMC and SCA is proposed as shown in Algorithm $2$ and Algorithm $3$ respectively. The computational complexity of BKMC and SCA is $O(K^3)$ and $O(K)$ respectively. Moreover, after obtaining the optimal resource allocation, we execute PPO-DRL as mentioned in Algorithm $3$ to get the optimal UAVs trajectories policies and its complexity is $O(Ka^2)$, where $a$ represents the number of actions. It can be observed that the proposed algorithms converged to the sub-optimal solutions.
\begin{algorithm}[t!]
	\caption{\strut PPO-DRL for UAVs Trajectory Optimization (\ref{opt:D})} 
	\label{alg:PPO}
	\begin{algorithmic}[1]
		\FOR{episode$=1,2,...,E$}
		\STATE{Initialize randomly each GU's positions}
		\STATE{GUs Association $\boldsymbol{\mathcal{\alpha}}$ by \textbf{Algorithm 1}}
		\FOR{actor$=1,2,...,A$}
		\FOR{time slot$=1,2,...,N$}
		\STATE{Run policy $\pi_{\theta_{\textrm{old}}}$ in environment}
		\STATE{Optimal Power Allocation $\boldsymbol{\mathcal{P}}$ by \textbf{Algorithm 2}}
		\STATE{Save $(s_{n},a_{n},r_{n},s_{n+1})$ in Trajectory memory}
		\ENDFOR
		\STATE{Compute advantage estimates $\hat{A}_{1},...,\hat{A}_{N}$}
		\ENDFOR
		\STATE{Optimize surrogate $L^{\textrm{PPO}}$ wrt $\theta$, with minibatch from Trajectory memory}
		\STATE{$\theta_{\textrm{old}} \leftarrow \theta$}
		\ENDFOR
		\STATE{\textbf{Output:} The optimal PPO network $\pi_{\theta_{\textrm{opt}}}$}
	\end{algorithmic}
	\label{alo3}
\end{algorithm}
\section{Simulation Results}
\label{simul}
In our network configuration, we consider $3$ UAVs and $36$ GUs distributed by following  homogeneous PPP within a $200$~m $\times$ $200$~m area in each episode. Moreover, UAVs are assumed to be hovering at a fixed altitude of $z_k$=$20$~m. The maximum speed of the UAVs is $5$~m/s. Other parameters are listed in Table I. All statistical findings are averaged after several simulation runs. To assess the performance of our proposed algorithm, we consider four benchmark algorithms as follows:
\begin{itemize}
    \item \textbf{SU with RP}: The algorithm which considers static UAVs (SU) positions with the random power (RP) allocation.
    \item \textbf{OU with RP}: The algorithm uses the optimal UAVs (OU) trajectory with the random power (RP) allocation.
    \item \textbf{SU with PP}: The algorithm assumes the static UAVs (SU) positions with the proposed power (PP) allocation.
    \item \textbf{OU with PP (proposed method)}: The algorithm considers the optimal UAV (OU) trajectory with the proposed power (PP) allocation.
\end{itemize}
\setlength{\arrayrulewidth}{0.15mm}
\setlength{\tabcolsep}{1.8pt}
\renewcommand{\arraystretch}{1}
\begin{table}[t]
\centering
\caption{Simulation Parameters}
\label{sim_tab}
\scalebox{1}{
\begin{tabular}{|c|c|c|c|}
\hline
    \textbf{Parameter}& \textbf{Value} & \textbf{Parameter}& \textbf{Value}\\ \hline \hline
    Bandwidth & $B$=$0.1$~THz & Channel gain at ref. & $h_{0}$=$-40$~dBm\\ \hline
	Noise power  & $\sigma^{2}$=$-174$~dBm/Hz  &  Max. transmit power & $P^{\textrm{max}}$=$2$~W\\ \hline
    Minimum rate & $R^{\textrm{min}}$=$0.02$~Tbps  & Absorption coefficient  & $a(f)$=$0.005$ \\ \hline
	Episodes   & $E$=$5e+5$ & Batch size   & $120$\\ \hline
	Discount factor & $\gamma~$=$0.99$ & Learning rate & $0.0003$  \\ \hline
	Clipping $\epsilon$ & $0.2$ & Regularizer parameter & $\lambda=0.95$\\ \hline
	Epochs & $3$ & Hidden layer's units& $128$ \\ \hline
	Hidden layers & $2$  & Carrier Frequency & f=$1.2$~THz \cite{THz_channel_ref}\\ \hline
	\end{tabular}}
\end{table}

\begin{figure}[t]
\centering
\includegraphics[width=\columnwidth]{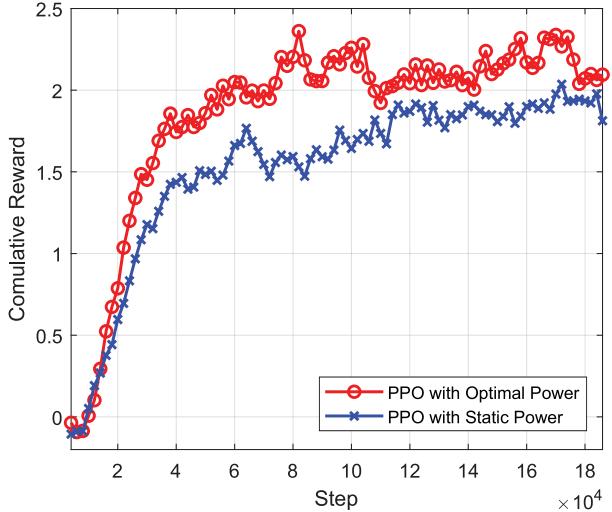} 
\caption{PPO learning results (reward).}
\label{convergence}
\end{figure}
Fig. \ref{convergence} shows the convergence performance of the cumulative reward with two power allocation schemes, i.e., optimal and static. Initially, both schemes provide fewer rewards, but after the convergence, the proposed schemes provide better results than the static schemes. 

\begin{figure}[t]
\centering
\includegraphics[width=\columnwidth]{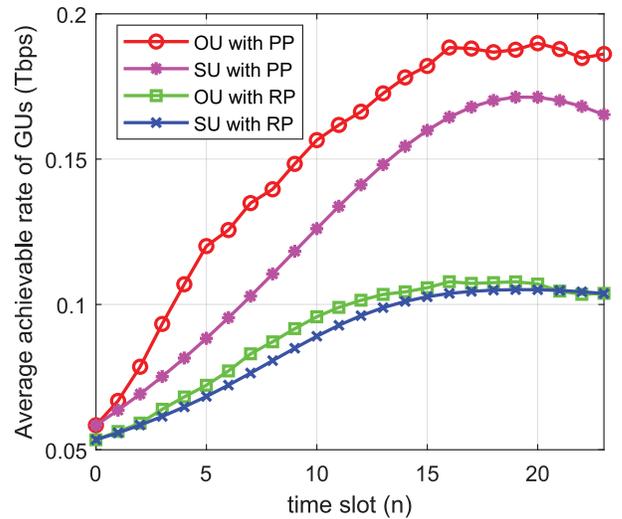}  
\caption{Achievable rate with benchmarks schemes.}
\label{benchmark}
\end{figure}
Fig. \ref{benchmark} provides a comparison of the proposed algorithm (OU with PP) with three benchmarks, i.e., SU with RP, OU with RP, and SU with PP in terms of the average achievable rate for GUs. It can be observed that OU with PP outperforms the rest of the three algorithms just after a few time slots. 

\begin{figure}[t]
\centering
\includegraphics[width=\columnwidth]{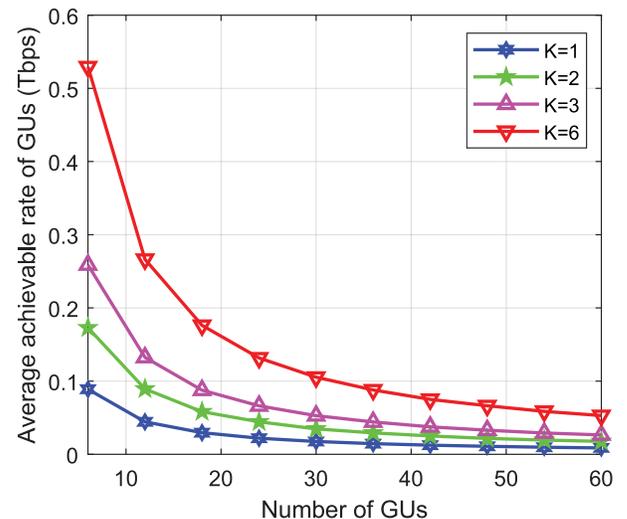}  
\caption{Achievable rate with UAVs.}
\label{comparison}
\end{figure}
Moreover, Fig. \ref{comparison} depicts the average achievable rate of GUs by varying the number of deployed UAVs in the area. Since all UAVs need to fulfill the QoS of GUs, and therefore, as the number of GUs increases in the area, network performance degrades. But, it can be observed that as the number of UAVs increases, the average performance increases.
\begin{figure}[t]
    \centering
    \includegraphics[width=\columnwidth]{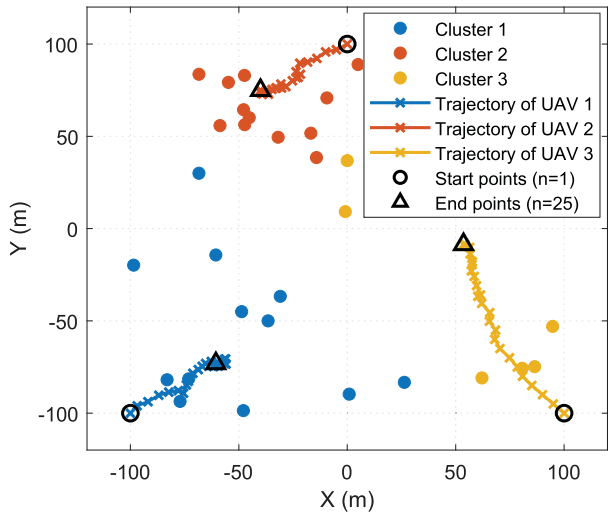}
    \caption{UAVs trajectory obtained by PPO-DRL.}
    \label{trajectory}
\end{figure}

The resultant UAV trajectory with the proposed power and GU allocation algorithm is shown in Fig. \ref{trajectory}. It is clear that the dots represent the GUs, and their clusters are separated by distinct colors. Each UAV follows the PPO-DRL algorithm to optimize its trajectory from an initialized location to the endpoint in time slot $n=1$ to $n=25$. Each UAV follows a distinct trajectory while providing optimal network resources to the GU.
\section{Conclusions}
\label{conclusion}
In this article, we have explored THz-enabled UAVs to facilitate ubiquitous 6G mobile communication networks. The molecular absorption effect has been explicitly incorporated in the THz-enabled UAV channel gain model. Then, we have formulated an optimization problem to optimize the average throughput of deployed UAVs by enhancing UAV-GU association, transmit power, and trajectories while satisfying the GU's demands. To address this problem, we have proposed an iterative algorithm that separates the original problem into three subproblems. Firstly, to tackle the UAVs-GUs association problem, we have employed the BKMC algorithm. To deal with the optimal transmit power, we have utilized the SCA-based algorithm. To handle dynamic UAV trajectories optimization, a PPO-DRL-based algorithm is has been designed, which can make quick decisions in the given environment owing to its low complexity. Based on the experience replay and target networks, the PPO method has efficiently learned the optimal trajectory with fast convergence speed. The simulation results have shown that our proposed algorithms outperform the other baselines.
\ifCLASSOPTIONcaptionsoff
\newpage
\fi
\bibliographystyle{IEEEtran}
\bibliography{ICC}
\end{document}